\input{aipcheck}

\documentclass[
    ,final            
  ]
  {aipproc}
\usepackage{epsfig,lscape}
\usepackage{bm}

\newcommand{\tauiso}{{\mbox{\boldmath $\tau$}}}

\layoutstyle{6x9}

\begin{document}

\title{Baryon Interactions in Nuclear and Hypernuclear Matter}

\author{H. Lenske, C. Keil, R.Shyam}{
  address={Institut f\"ur Theoretische Physik, Universit\"at Giessen, D-35392
  Giessen}
  ,altaddress={$^\dag$Saha Institute of Nuclear Physics,
Calcutta, India} }
%
%

\begin{abstract}

The production and structure of $\Lambda$ hypernuclei are
investigated in field theoretical models. The production in
coherent p+A reactions is investigated by means of a resonance
model. Results of exploratory calculations for associated
strangeness production are presented for proton reactions on
$^{40}$Ca. The target single particle wave functions are obtained
from DDRH theory with density dependent Dirac-Brueckner
meson-baryon vertices. The dependence of the in-medium vertices on
density is discussed.
\end{abstract}
\pacs{ $25.40.Ve$, $13.75.-n,$, $13.75.Jz$ }

\maketitle


\section{Introduction}

A major goal of hadron physics is to understand the interactions
among the various members of the baryonic and mesonic flavor
multiplets. For that aim an important class of processes is the
implementation of a strange hadron into a nuclear environment by
means of an appropriate reaction and investigations of the
subsequent evolution of that system. Here we consider interaction
within the lowest baryon flavor octet. We are especially
interested in associated $K^+\Lambda$ strangeness production in
hadronic reactions. As an interesting alternative to more common
approaches using pion and kaon reactions \cite{Chrien:89} we
consider strangeness production in exclusive $p+A \to
_\Lambda(A+1)+K^+$ reactions at incident kinetic energies in the
COSY energy regime. A field theoretical model is used describing
associated strangeness production by the excitation of
intermediate nucleon resonances decaying into the $K^+\Lambda$
channel. Since we are aiming at hypernuclei only those processes
in which the hyperon in captured in a bound orbit are considered.
Clearly, since these are highly selective reactions at large
momentum transfer, a high sensitivity on nuclear wave functions
has to be expected. We use DDRH theory \cite{Lenske:95} as a
state-of-the-art relativistic field theory for the nuclear
structure calculations, known to describe both normal nuclei
\cite{Hofmann:01} and hypernuclei \cite{Keil:00} very accurately.
This allows us to investigate the production cross sections for
various nuclear states thus establishing the link from the
production to the spectroscopy of the final hypernuclei and
sampling their wave function at momenta which otherwise are not
accessible.

\section{Associated Strangeness Production in Proton-Nucleus
Reactions}\label{sec:production}

The production of $\Lambda$-hypernuclei with high intensity proton
beams is an interesting alternative to more common pion and
antikaon induced production scenarios \cite{Chrien:89} . In
principle, the production process can proceed in a variety of
reactions like $p + A(N,Z) \rightarrow {_{\Lambda}}B(N-1,Z) + n +
K^+$, $p + A(N,Z) \rightarrow {_{\Lambda}}B(N,Z-1) + p^\prime +
K^+$, and $p + A(N,Z) \rightarrow {_{\Lambda}}B(N,Z) + K^+$ where
$N$ and $Z$ are the neutron and proton numbers, respectively, in
the target nucleus. Here, we study the last reaction [to be
referred to as $A(p,K^+){_{\Lambda}}B$] which is exclusive in the
sense that the final channel is a two body system. In this
reaction the momentum transfer to the nucleus is much larger than
in $(\pi^+,K^+)$ reaction, about 1.0 GeV/c as compared to about
0.330 GeV/c in forward direction.

The elementary production process is a two-nucleon mechanism (TNM)
\cite{Shyam:03,Shyam:99} where the kaon production proceeds via a
collision of the projectile nucleon with one of its target
counterparts, thereby exciting intermediate baryonic resonances
decaying in turn into a kaon and a $\Lambda$ hyperon. The
$N^*1650)$, $N^*1710)$, and $N^*1720)$ states are espcially
important \cite{Shyam:03}. The nucleon and the hyperon are
captured into nuclear orbitals while the kaon is rescattered onto
its mass shell. Three active bound state baryon wave functions are
taking part in the reaction process allowing the large momentum
transfer to be shared among the participants.

We use a field theoretical approach with effective Lagrangians for
the nucleon-nucleon-pion ($NN\pi$) and $N^*$-nucleon-pion
($N^*N\pi$) vertices \cite{Shyam:03}. They are given by
\begin{eqnarray}
{\cal L}_{NN\pi} & = & -\frac{g_{NN\pi}}{2m_N} {\bar{\Psi}} \gamma
_5
                             {\gamma}_{\mu} \tauiso
                            \cdot (\partial ^\mu \boldmath{\Phi}_\pi) \Psi. \\
{\cal L}_{N_{1/2}^*N\pi} & = & -g_{N_{1/2}^*N\pi}
                          {\bar{\Psi}}_{N_{1/2}^*} i{\Gamma} \tauiso
                        \boldmath{\Phi}_\pi \Psi
                          + {\rm h.c.},\\
{\cal L}_{N_{3/2}^*N\pi} & = & \frac{g_{N_{3/2}^*N\pi}}{m_\pi}
                         {\bar{\Psi}}_{\mu}^{N^*} \Gamma_\pi
                         {\tauiso} \cdot \partial ^{\mu}
                         \boldmath{\Phi}_\pi \Psi + {\rm h.c.}.
\end{eqnarray}
where $m_N$ denotes the nucleon mass. The operator $\Gamma$
($\Gamma_\pi$) is either $\gamma_5$ (unity) or unity ($\gamma_5$)
depending upon the parity of the resonance being even or odd,
respectively. Following Ref.~\cite{Shyam:99} we use a pseudovector
(PV) coupling for the $NN\pi$ vertex and a pseudoscalar (PS) one
for the $N_{1/2}^*N\pi$ vertex. The effective Lagrangians for the
resonance-hyperon-kaon vertices are written as
\begin{eqnarray}
{\cal L}_{N_{1/2}^*\Lambda K^+} & = & -g_{N^*_{1/2}\Lambda K^+}
                          {\bar{\Psi}}_{N^*} {i\Gamma} \tauiso
                        \boldmath{ \Phi}_{K^{+}} \Psi
                        + {\rm h.c.}.\\
{\cal L}_{N_{3/2}^*\Lambda K^+} & = & \frac{g_{N^*_{3/2}\Lambda
K^+}}{m_{K^+}}
                         {\bar{\Psi}}_{\mu}^{N^*} \Gamma_\pi
                         {\tauiso} \cdot \partial ^{\mu}
                         \boldmath{ \Phi}_{K^{+}} \Psi + {\rm h.c.}.
\end{eqnarray}
Here, ${\Psi}_{\mu}^{N^*}$ is the vector spinor for the spin-${3
\over 2}$ particle. Further discussions about the vertices and
coupling constants involving such particles are found in
Refs.~\cite{Shyam:03,Shyam:99,Penner:02}. The amplitude for graph
1b with spin-${1 \over 2}$ baryonic resonance, for example, is
given by,
\begin{eqnarray}
M_{1b}(N^*_{1/2}) & =&
C_{iso}^{1b}\biggl(\frac{g_{NN\pi}}{2m_N}\biggr)
(g_{N_{1/2}^*N\pi}) (g_{N^*_{1/2}\Lambda K^+})
{\bar{\psi}}(p_2) \gamma _5 \gamma_\mu q^\mu \nonumber \\
& \times & \psi(p_1) D_{\pi}(q)
{\bar{\psi}}(p_\Lambda)\gamma_5 D_{N^*_{1/2}}(p_{N^*}) \gamma_5 \nonumber \\
& \times & \phi^{(-)*}_{K}(p_K^\prime,
p_K)\psi^{(+)}_i(p_i^\prime, p_i),
\end{eqnarray}
where various momenta are as defined in Fig.~1b. For a more
detailed discussion we refer to ref. \cite{Shyam:03}.

\begin{figure}
\centering \epsfig{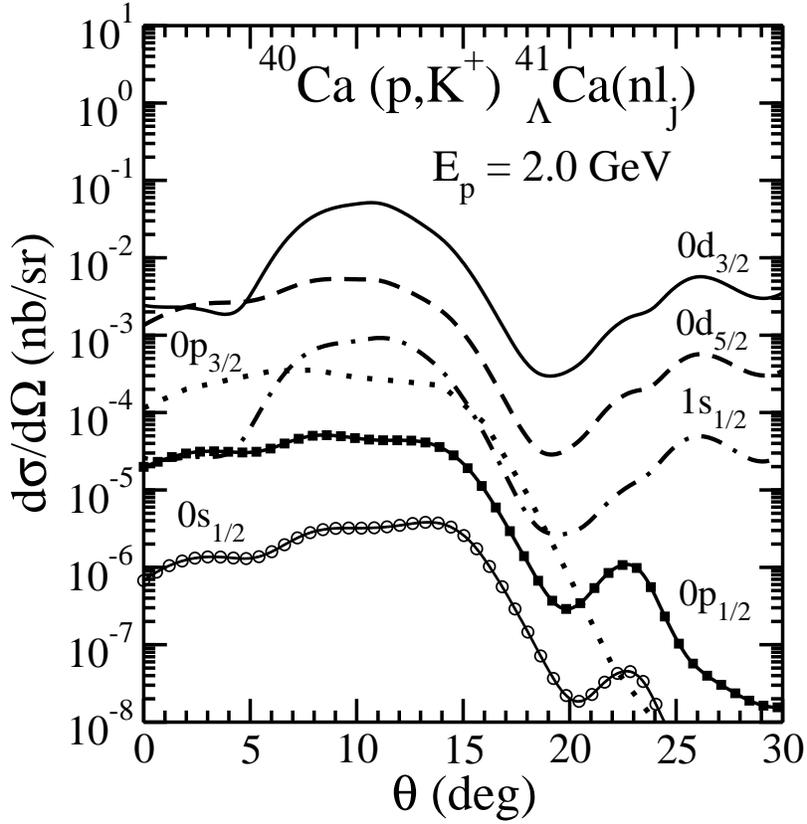}
\caption{Angular distributions for production into various final
$\Lambda$ single particle orbitals.}
\label{fig:Orbits}       
\end{figure}

Angular distributions for associated strangeness production in a
$(p,K^+)$ reaction at E$_p=$2~GeV on a $^{40}$Ca target are shown
in fig. \ref{fig:Orbits}. Although initial and final state
interactions are at present not included the magnitude of the
cross sections in the nano- to picobarn range can be expected to
be realistic, ranging at the lower end of the experimental
feasibility. The shapes of the angular distributions are depending
sensitively on the quantum numbers of the orbits into which the
$\Lambda$ is captured.

\section{Interactions in Hypermatter and Hypernuclei}\label{sec:spectros}

The nuclear wave functions entering into the cross section
calculation have been obtained by DDRH theory. This is a field
theoretical approach accounting for the modifications of
baryon-baryon interactions in matter with density dependent
meson-baryon vertices. In \cite{Lenske:95} it was shown that a
properly defined field theory, preserving covariance of the field
equations and thermodynamical consistency, is obtained when vertex
functionals depending on the field operators are used. The medium
dependence of the vertices is derived from Dirac-Brueckner (DB)
calculations and applied in relativistic mean-field calculations
to nuclei \cite{Lenske:95,Hofmann:01} and hypernuclei
\cite{Keil:00,Keil:02}. Hence, we obtain an {\it ab initio}
description once the free space baryon-baryon interaction is
specified. In the strangeness sector, however, the information is
sparse making it necessary to introduce the ratio of the scalar to
the vector meson coupling constant as a free parameter which is
determined from hypernuclear spectra \cite{Keil:00}.
Interestingly, we find significant deviations from the
expectations of the naive quark model: While the latter predicts a
reduction of the meson-$\Lambda$ vertices by $1/3$
\cite{Chrien:89} the analysis of the existing data on $\Lambda$
hypernuclei imply a reduction factors of about 50-60\%
\cite{Keil:00}. The single $\Lambda$ separation energies are well
described by our calculations, especially for the heavier mass
region, $A>40$. Discrepancies for low masses, $A < 16$, seem to be
related to additional dynamical self-energies coming from core
polarization, thus supporting a conjecture of Polls et al.
\cite{Polls:98}

\begin{figure}
\centering
\epsfig{file={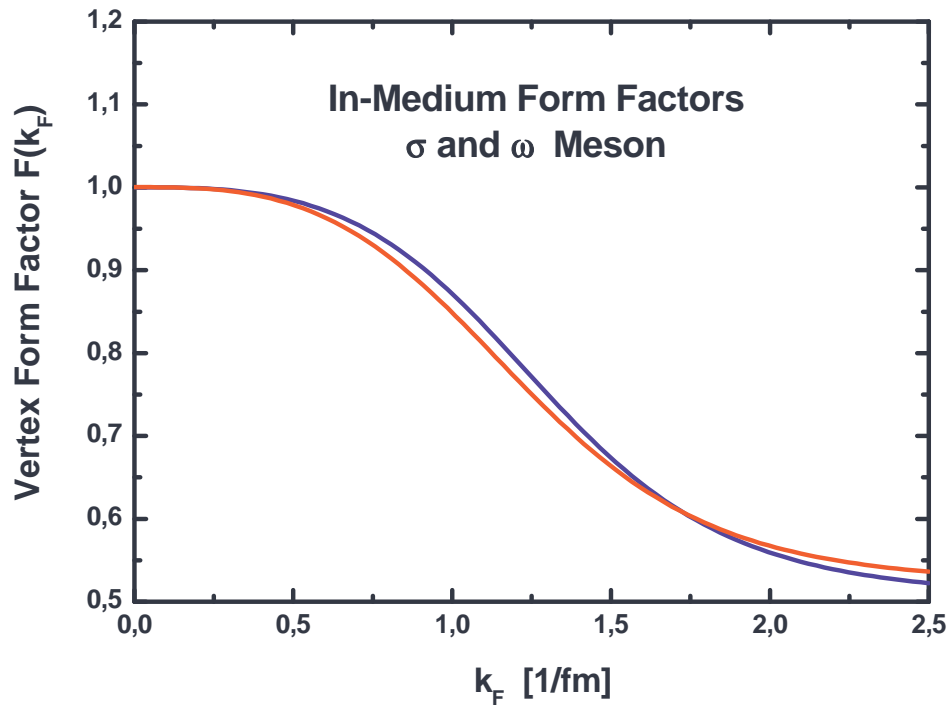},width=0.85\textwidth}
\caption{In-medium vertex form factors in the isoscalar meson
channels describing the variation of the meson-baryon coupling
strength with density.}
\label{fig:FormF}       
\end{figure}

It is worthwhile to consider more closely the general properties
of in-medium interactions for the $SU(3)_f$ flavor octet baryons.
Denoting the in-medium vertices for meson $\alpha=\sigma,\omega$
by $\Gamma_{\alpha B}(\rho)$ the ratio of $\Lambda$ to nucleon
coupling constants is found to behave as \cite{Keil:00}
\begin{equation}
R_\alpha = \frac{\Gamma_{\alpha\Lambda}}{\Gamma_{\alpha N}} =
\frac{g_{\alpha\Lambda}}{g_{\alpha
N}}+\mathcal{O}\left((\frac{k^{\Lambda}_F}{k^{N}_F})^2\right)
\end{equation}
showing that the in-medium vertices indeed reflect in leading
order the free space properties of the couplings. The above
relation indicates that the density dependence of baryons is most
likely given by a common form factor depending on the Fermi
momentum $k^q_F$ of the flavor species $q$. In fact, inspection of
the in-medium Bethe-Salpeter equation shows that medium
dependencies are introduced primarily through Pauli-blocking which
obviously is related to particles of the same flavor
\cite{Lenske:03}. As an overall effect, the intrinsic density
dependence of the meson-baryon vertices leads to a considerable
variation of the coupling strength over the nuclear volume
\cite{Keil:00,Keil:02}, suppressing the coupling with increasing
density. In fact, the in-medium interactions can be represented in
terms of a susceptibility tensor and the free space interaction,
at least in the ladder approximation \cite{Lenske:03}. In the DB
vertices $\Gamma_{\alpha q}(k^q_F)=G_{\alpha q} F_{\alpha
q}(k^q_F)$, obtained by solving the BS equation in ladder
approximation, we may express the density dependence by form
factors $F_{\alpha q}(k^q_F)$ and an overall strength $G_{\alpha
q}$. From fig. \ref{fig:FormF} it is seen that the shape of the
form factor is almost independent of the meson channel thus
indicating universality.

\section{Summary and Outlook}

The formation of hypernuclei in exclusive proton-nucleus reactions
by associate ($K^+\Lambda$) strangeness production has been
discussed. The elementary vertices are described in a field
theoretical model assuming a two-step type process where initially
a nucleon is excited into a resonant state which subsequently
decays into the $K^+\Lambda$ final state and the hyperon is
captured by the target nucleus. Angular distributions for the
outgoing $K^+$ were found to depend rather sensitively on the
orbit occupied by the $\Lambda$. Hypernuclear structure was
described by DDRH theory. A field theoretical approach to the
density dependence of meson-baryon vertex functionals was
discussed.


\begin{theacknowledgments}
This work was supported in part by FZ J\"ulich, grant COSY-066.
\end{theacknowledgments}


\end{document}